\documentclass[a4paper,fleqn,usenatbib]{mnras}

\usepackage{newtxtext,newtxmath}

\usepackage[T1]{fontenc}
\usepackage{ae,aecompl}

\usepackage{float}

\usepackage{graphicx}	
\usepackage{amsmath}	
\usepackage{amssymb}	

\title{Retrograde orbits excess among observable interstellar objects}

\author[D. Mar\v ceta and B. Novakovi\' c]{
Du\v san Mar\v ceta,\thanks{E-mail: dmarceta@matf.bg.ac.rs (DM)}
Bojan Novakovi\'c
\\
Department of Astronomy, Faculty of Mathematics, University of Belgrade, Studentski trg 16, 11000 Belgrade, Serbia\\
}

\date{Accepted XXX. Received YYY; in original form ZZZ}

\pubyear{2019}

\begin{document}
\label{firstpage}
\pagerange{\pageref{firstpage}--\pageref{lastpage}}
\maketitle

\begin{abstract}
In this work we investigate the orbital distribution of interstellar objects (ISOs), observable by the future wide-field National Science Foundation Vera C. Rubin Observatory (VRO). We generate synthetic population of ISOs and simulate their ephemerides over a period of 10 years, in order to select those which may be observed by the VRO, based on the nominal characteristics of this survey. We find that the population of the observable ISOs should be significantly biased in favor of retrograde objects. The intensity of this bias is correlated with the slope of the size-frequency distribution (SFD) of the population, as well as with the perihelion distances. Steeper SFD slopes lead to an increased fraction of the retrograde orbits, and also of the median orbital inclination. On the other hand, larger perihelion distances result in more symmetric distribution of orbital inclinations. We believe that this is a result of Holetschek’s effects, which is already suggested to cause observational bias in orbital distribution of long period comets. The most important implication of our findings is that an excess of retrograde orbits depends on the sizes and the perihelion distances. Therefore, the prograde/retrograde orbits ratio and the median inclination of the discovered population could, in turn, be used to estimate the SFD of the underlying true population of ISOs.
\end{abstract}




\section{Introduction}
\label{sec:intro}

The existence of galactic population of the objects ejected from the planetary systems has been long hypothesized \citep[e.g.][]{Sekanina}. The expelling of a large number of planetesimals during the early stages of the Solar System is predicted by its evolution models \citep[e.g.][]{Charnoz2003, Bottke2005, 2011Natur.475..206W}, and is reasonable to assume that this process is also at work in other planetary systems throughout the Galaxy. Some authors claim that ejections in the early phase is not sufficient to match their estimated number density, and proposed other ejection mechanisms, including ejection of the planetesimals during the late phases of the stellar evolution process \citep{Veras2014, Stone2015}. The discovery of 1I/(2017 U1) 'Oumuamua, the first macroscopic interstellar object (ISO) by Pan-STARRS survey \citep{MPC-oumuamua}, not only confirmed their existence, but also indicated that the population of these objects is relatively numerous. In turn, as discussed by \citet{2018ApJ...855L..10D}, this enabled setting better constraints on their number density and size-frequency distribution (SFD). This is further supported by more recent discovery of the object 2I/(2019 Q4) Borisov \citep{MPC-borisov}, which is also confirmed to have the interstellar origin \citep{2019ApJ...886L..29J,2020A&A...634A..14B}.

One can say that 'Oumuamua was the exact opposite of what we expected from an interstellar object. This is primarily related to its extremely elongated shape and asteroidal nature. The estimates of its aspect ratio go from 3.5:1 \citep{2018ApJ...852L...2B} to 10:1 \citep{2017Natur.552..378M}. Although, there are small objects with comparable aspects ratios in the Solar System, such as asteroid (1865) Cerberus, whose aspect ratio is estimated to 4.5:1 \citep{2012A&A...547A..10D}, they are generally rare. Therefore, highly elongated shape of the very first known interstellar object 'Oumuamua, was highly unexpected. 
 
On the other hand, although models of planetary systems evolution predict that the large number of planetesimals should escape their mother systems, it is expected that large majority of these object should originate from the outer parts of the systems, far beyond the snow-line \citep{2018ApJ...852L..15C}. Hence, it was reasonable to expect that ISOs show cometary activity close to the perihelion. Although coma around 'Oumuamua was not detected directly, astrometric measurements showed deviation from a purely gravity driven trajectory, which may be explained with an additional force induced by cometary activity \citep{2018Natur.559..223M}. However, \citet{2018ApJ...867L..17R} argues that this amount of activity should have led to significant evolution of the object's rotational state, and probably to its disruption, but no significant evolution of the light curve was observed during this period. Unlike the latter study, \citet{2019ApJ...876L..26S} suggest that out-gassing activity that followed the sub-solar point of an elongated body could produce the observed non-gravitational acceleration, without causing extreme spin up. This, and many other questions about the 'Oumuamua, are still open \citep{2019NatAs...3..594O}.

The lack of observed typical cometary activity was not only surprising because of the disagreement with an expected nature of a vast majority of ISOs, but also because the probability of their discovery should be significantly biased in favor of cometary-like objects, due to increased brightness caused by the sublimation of volatile materials. Still, the second ISO (2I/Borisov) shows cometary activity, suggesting that we should expect a large variety of characteristics among ISOs, which will hopefully be discovered in the near future, especially after the start of the National Science Foundation Vera C. Rubin Observatory’s (VRO) Legacy Survey of Space and Time (LSST)\footnote{Formerly known as the Large Synoptic Survey Telescope (LSST)}.

Recent studies about ISOs number density and number of objects expected to be detected by the current and future surveys give large variety of results. A comprehensive analysis of ISOs number density by \citet{2009ApJ...704..733M} indicated that the probability for the VRO to detect an ISO during its operating period is very small, on the order of 0.001-1 \%. This result is based on a consideration of the expected ISOs number density, which included the number density of stars, the amount of solids available to form planetesimals, the frequency of planets and planetesimals formation, the efficiency of planetesimals  ejection, and the possible size distribution of these small bodies. However, the analysis was limited only to the ISOs orbiting beyond the orbit of Jupiter, and did not take into account a possibility that ISOs become active when approach closer to the Sun, that may significantly increase their brightness, and therefore chances to be detected. \citet{2016ApJ...825...51C} extended this analysis by taking into account gravitational focusing by the Sun (which increases the number of ISOs per unit volume closer to the Sun), the effect of different observing angles (photometric phase functions), comet brightening, and more precise definition of the observing constraints (such as solar elongation and air mass). These improvements allowed consideration of the detection of closer ISOs, leading to an estimation of 0.001 to 10 expected detections of ISOs by the VRO during the 10 years of its nominal operating period.

Such a small number of expected detections is mainly a consequence of the estimated number density of ISOs. However, \citet{2017AJ....153..133E} determined the upper limit for the ISOs number density to be several orders of magnitudes larger than previously estimated. Their analysis is based on a modeling of ISOs population around the Sun, that naturally includes the effect of gravitational focusing. The authors exposed this population to detectability simulation based on the performances of three surveys (Pan-STARRS1, Mt. Lemmon Survey, and Catalina Sky Survey), and considered the different effects, including cometary activity, photometric phase functions, observing constellations, and various SFD functions. In addition, \citet{2017AJ....153..133E} based their findings on the fact that no single ISO was discovered at that time. Therefore, the recent discoveries of 'Oumuamua and Borisov, suggest that it may not be a surprise if the VRO detects even larger number of ISOs, than expected in the most optimistic predictions \citep[see][]{2019Sci...366..558G}. 

While a nominal number of the detectable ISOs is definitely an important parameter to know, the observational selection effects may play important role in analyzing and modeling the underlying populations \citep{2002aste.book...71J}. Still, many aspects of the observational selection effects on ISOs population have received very little attention in the literature so far. The goal of the work presented in this paper is twofold: i) to determine the orbit and size-frequency distribution of the ISOs observable by the VRO, and ii) to analyze how these distributions depend on the same properties of the underlying true population.

\section{The population of interstellar objects}
\label{sec:population}

In order to perform the analysis, it is necessary to define some input parameters, make some assumptions and adopt some methodologies. Below we outline our approach.

\subsection{Number density and size distribution of ISOs}
\label{ss:num-density}

A total number of objects which can be detected by an observation program primarily depends on how many of them are in the observable volume of the space, and how large (bright) they are. Hence, the two most important parameters that determine the detection probability of ISOs are their number density and SFD. However, due to the lack of observational data, the estimations of these parameters are based primarily on theoretical assumptions and, consequently, are very uncertain. 

There is a large dispersion of the assumptions for the ISOs number density, and for objects larger than 1 km in diameter, it ranges from $10^{-9}$ au$^{-3}$ \citep{2009ApJ...704..733M} to $10^{-2}$ au$^{-3}$ \citep{2017AJ....153..133E}. This is a consequence of limited knowledge about how efficiently planetary systems populate the interstellar space with asteroids and comets. A number of ejections in the early phases depends on various characteristics of the systems, such are their orbit architectures, or masses of the planets. In addition, as mentioned before, it is possible that other mechanisms characteristic for the late phases of planetary systems evolution also contribute to this process.

A similar situation is also with the SFD of ISOs. For instance, it is unknown if it represents their initial population, as they were expelled from their mother planetary systems, or it is significantly altered during their interstellar phase. These concerns naturally arise from the attempts to explain the lack of cometary activity and extremely elongated shape of 'Oumuamua. As an example, \citet{2019MNRAS.484L..75V} suggest that the elongated shape is a consequence of isotropic erosion. If true, this should also shrinks sizes of ISOs, significantly altering their SFD, to such an extent that it may be even responsible for lack of the observable objects.

The main goal of this work is not to estimate the exact number of objects which will be detected and eventually discovered by the current and future survey programs, but to analyze the orbital and size distributions of the detectable objects. To this purpose, we assumed that a cumulative size-frequency distribution of ISOs is given in the standard, single slope power-law form $ N(>D)\propto D^{-\gamma}$, where $D$ and $\gamma$ are objects' diameters and SFD slope, respectively. The analyzes were then performed assuming the range of SFD slopes $\gamma$ between $1.4$ and $4$, with a discrete step of $0.1$.  A population of main-belt asteroids, larger than about 2~km in diameter, has a cumulative size-frequency distribution characterized by a slope of $\gamma = 2.4$ \citep{2015A&A...578A..42R}. For Jupiter Family Comets (JFC), \citet{2013Icar..226.1138F} found a shallow $\gamma$ slope of $1.9$ in the size range $2.8 - 18$~km, while for Long Period Comets (LPCs), \citet{2019Icar..333..252B} found a slope of $\gamma = 3.6$ for objects larger than $1$~km. Therefore, although the interval of slopes analyzed here is selected somewhat arbitrary, it covers the values available in the literature for possibly representative populations, and even extends for about 0.5 on both sides with respect to the interval of quoted slopes.\footnote{We note that, generally, it seems that population of the small Solar System objects has a shallower SFD at small than at larger sizes \citep{2009Icar..202..104G,2014Icar..231..168B, 2019Sci...363..955S}, and therefore should be represented as a broken power-law. In this work we did not consider these findings, but it would be worth to model also populations with
broken power-law slopes in the future work.} The range of analyzed slopes allows to highlight any possible connection between the orbit distribution and the SFD.

Furthermore, we generated the population with a number density of $10^{-4}$ au$^{-3}$ objects larger than 1 km in diameter, which is between the extremes of the previous assumptions \citep{2009ApJ...704..733M,2017AJ....153..133E}. For the SFD slope of 2.5, which is expected for the so-called self-similar collisional cascade \citep{1969JGR....74.2531D}, this number density corresponds to 10 ISOs per au$^3$ larger than 10 meters in diameter. While the number density is the crucial parameter for the estimation of the absolute number of objects which could be observed, it is not expected that this parameter influence the orbital and size distributions of the observable objects, because it equally impacts all objects from the population, regardless of their sizes and orbits. Having this in mind, we chose the number density that is within the bounds of the previous estimates, can be treated with available computing resources, and provides a sufficient sample for statistical analysis.

\subsection{Orbital elements of ISOs}
\label{ss:orb_ele}

It seems appropriate to suppose that the population of ISOs throughout the Galaxy, far from any massive body, is homogeneous, and that their velocity vectors are isotropic. Also, it is reasonable to assume that the distribution of their speeds mimic that of the nearby stars. However, in the area close to the Sun, these distributions will be altered due to the effect of gravitational focusing. In order to generate the steady-state population of ISOs in the vicinity of the Sun, we applied a modified method of \citet{2011PASP..123..423G}. 

In particular, for the purpose of the analyses performed in this work, we use
a concept of three spheres: observable, model and initialization \citep[see also][]{2017AJ....153..133E}. The idea behind this concept is the following. We are interested in the ISOs potentially observable by the VRO. Therefore, we define the radius of the {\it observable sphere} in such a way that at least brightest objects from the population should be visible at the edge of this sphere. However, in order to determine the population of the ISOs situated inside the observable sphere, we need to model the population in a larger volume of space, from which objects can enter the observable sphere during the 10 yr operational period of the VRO. This sphere that should feed the observable sphere is called {\it model sphere}. The radius of the model sphere should be large enough to include all the objects that can reach the observable sphere. Therefore, it is defined based on the distance that the fastest objects will cross in 10 yr. The distribution of the objects inside the model sphere is not uniform, due to the gravitational focusing in vicinity of the Sun. For this reason, we need to define an additional volume of space that in turn feeds the model sphere, but which is far enough from the Sun that the gravitational focusing may be neglected. This is what we called {\it initialization sphere.}

A detailed technical description of our methodology is given below.
\begin{enumerate}
\item We used simulation time of 10 years, that is nominal operating period for the VRO. This means that our population should have unchanging characteristics, at least, within this period.
  
\item We set limiting apparent visual magnitude of $m = 24.5$, which is the nominal limiting magnitude of the VRO \citep{2016IAUS..318..282J}. This practically means that the brightest object from the population can reach this limiting magnitude at the edge of the observable space, under ideal observing conditions.
 
\item We worked with objects between $10$~m and $10$~km in diameter. Based on the predictions of ISOs number density and SFD, there should be comparatively few ISOs larger than $10$~km, and it is therefore unlikely they will penetrate the inner Solar System. For the most optimistic assumptions of the number density \citep[e.g.][]{2017AJ....153..133E} and moderate SFD slopes, we can expect on the order of $10^{-1}$ ISOs larger than $10$~km, inside the observable volume of space at any instance. On the other hand, an object of $10$~m in size has to pass within $0.1$~au from the Earth, while being in the opposition, in order to reach the limiting apparent magnitude. For this reason, it is highly unlikely that objects smaller than $10$~m will be detected, due to their faintness, but also extremely large apparent velocities if they appear close enough to the Earth.

\item We assumed that the distribution of speeds of these objects relative to the Sun, when they are at infinity (hyperbolic excess velocity, $v_{\infty}$), mimics the distribution of speeds of the nearby stars. We adopted the normal distribution with mean value of $v_0 = 25$ km/s, and standard deviation of $\sigma=5$ km/s \citep[e.g.][]{1998MNRAS.298..387D}. This means that 99.75$\%$ of the objects have speeds between $10$ and $40$~km/s.

\item The relation between diameter and absolute magnitude is calculated according to the following relation \citep{1997Icar..126..450H}: 
\begin{equation}
H=15.618-2.5\log\left(p_v\right)-5\log\left(D\right),
\label{eq:H(D)}
\end{equation}
where diameter ($D$) is given in kilometers and $p_v$ is geometric albedo, for which we adopted a value of 0.04 for the whole population, since it is the estimated value for 'Oumuamua \citep{2017Natur.552..378M}. The apparent magnitude of an ISO is calculated from the equation: 
\begin{equation}
   m = H+5log(r_g r_h)+\Phi \left(\alpha \right),
\label{eq:m(H)}
\end{equation}
where $m$ is apparent visual magnitude, $H$ is absolute magnitude, $r_g$ and $r_h$ are geocentric and heliocentric distances, respectively, and $\Phi \left(\alpha \right)$ is the integral phase function of the phase angle ($\alpha$) \citep{1989aste.conf..524B}. We adopted simplified linear darkening function $\Phi \left(\alpha \right)=\beta \alpha$ with a slope of $\beta = 0.04$ degree$^{-1}$, which neglects brightening due to opposition surge \citep{2017ApJ...850L..36J}. According to Eq.~\ref{eq:H(D)}, the brightest object from the generated population, with a diameter of 10 km, has absolute magnitude of $H = 14.1$. From Eq.~\ref{eq:m(H)}, we obtained the limiting heliocentric distance of $11.5$~au, at which the brightest object can reach the limiting apparent magnitude, when it is at the opposition ($\Phi = 0$; $r_g=r_h-1$~au). Based on this, we adopted the value of $12$~au as the radius of the sphere that represents the observable volume of space.

\item We calculated the heliocentric distance from which the fastest objects from the population ($v_{\infty} = 40$ km/s), assuming they are on direct paths toward the Sun, can reach the observable sphere during the 10 years of the simulation interval. This heliocentric distance is found to be at $97$~au, which we adopted for the radius of the model sphere.
  
\item As mentioned before, far enough from the Sun, it is reasonable to expect that population of ISOs is homogeneous and that their velocity vectors are isotropic. However, in vicinity of the Sun this assumption will be disturbed as a result of gravitational focusing. A parameter that determines the intensity of the gravitational focusing is defined as $F=1+ {{v_{esc}}^2 / {v_{\infty}}^2}$ \citep[see e.g.][]{2017ApJ...850L..36J}, where in our case $v_{esc}$ is the escape velocity from the Sun, at a given heliocentric distance. For the average hyperbolic excess velocity of the generated population ($v_{\infty} = 25$ km/s), a value of this parameter at the edge of the model sphere ($r_m = 97$ au, $v_{esc} = 4.28$ km/s) is only $F \approx 1.03$. Hence, we assumed that outside the model sphere population of ISOs is unaffected by the gravity of the Sun, and thus it is homogeneous and isotropic. To obtain the population inside the model sphere, where the gravitational focusing cannot be neglected, this space has to be populated only with the objects initially located outside it. The simulation of the process of populating has to last long enough, that the population inside the model sphere can reach approximately a constant number. To achieve this, all objects at the edge of the model sphere, should have enough time to cross this sphere, regardless of their initial velocity vectors. We called this time the {\it initialization time}. On the other hand, the size of the initialization sphere, that populates the model sphere, should coincide with the distance from which the fastest objects can reach the model sphere during the initialization time. 

To estimate the initialization time, we determined the longest time taken by an object from the population to cross the entire model sphere. This calculation is performed over the whole range
of values of $v_{\infty}$ and $q$. For $v_{\infty}$ this range goes from $10$ to $40$~km/s, while for the perihelion distances we considered the range of distances between the radius of the Sun ($0.005$~au) and $97$~au, which is the radius of the model sphere. These ranges of $v_{\infty}$ and $q$ were then sampled on equidistant grid points, and for any possible combination of these two parameters we calculated the semi-major axis ($a$) and eccentricity ($e$), according to the equations \citep{kemble_2006}:
\begin{equation}
    a = - {{\mu} \over {v_{\infty}^2}},
\label{eq:a(v)}
\end{equation}

\begin{equation}
    e=1-{{q} \over {a}}
\label{eq:e(q)}
\end{equation}
\noindent where $\mu$ is the gravitational parameter of the Sun. Having $a$ and $e$, a critical hyperbolic anomaly ($H_{cr}$) that corresponds to the edge of the model sphere, can be obtained from the equation for a hyperbolic orbit:
\begin{equation}
    r=a \left(1 - e\cosh{H} \right),
\label{eq:r(H)}
\end{equation}
where $H$ is hyperbolic anomaly. Finally, we calculated the time needed for an object to move along the hyperbolic trajectory from $-H_{cr}$ to $H_{cr}$ according to the hyperbolic Kepler equation
\begin{equation}
    M=e\sinh{H}-H,
\label{eq:H}
\end{equation}
where M is mean anomaly.

The obtained times taken by objects from our generated population to cross the model sphere are shown in Fig.~\ref{fig:time_spent}, as a function of perihelion distance, eccentricity and hyperbolic excess velocity. 
  
\begin{figure}
	\includegraphics[width=\columnwidth]{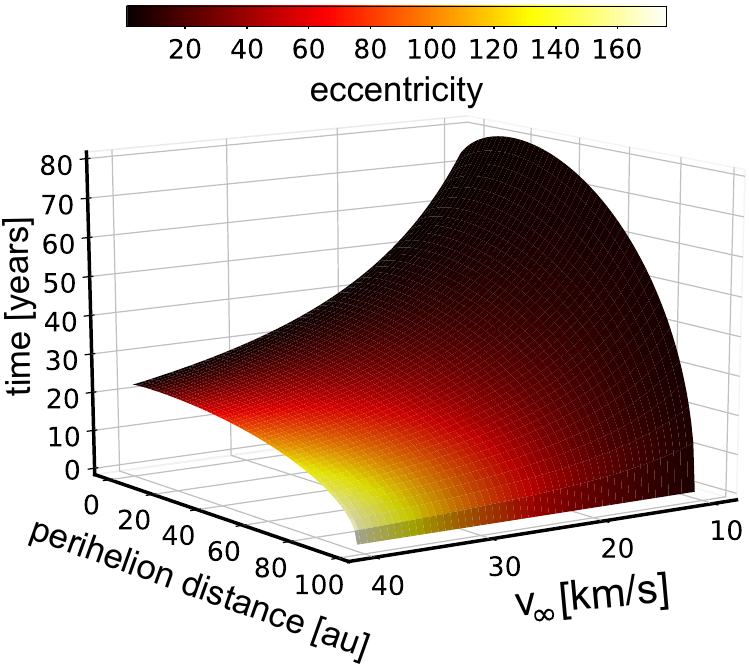}
    \caption{Surface shows the time that an interstellar object spends inside the model sphere, of $97$~au radius, depending on $v_{\infty}$ and perihelion distance. The color scale represents orbital eccentricity.}
    \label{fig:time_spent}
\end{figure}
We found that the longest time to cross the model sphere of 80 years needs an object with $v_{\infty} = 10$~km/s, $q = 21.45$~au and $e = 3.43$. Therefore, all objects inside $3\sigma$ limits of $v_{\infty}$, at the edge of the model sphere, have enough time to leave it in 80 years.

\item To calculate the radius of the initialization sphere, we calculated the distance from which the fastest object from the population ($v_{\infty} = 40$ km/s), on its direct path to the Sun, can reach the model sphere in 80 years, and we obtained the value of $800$~au. This means the model sphere in the time of 80 years will be almost entirely populated with the objects which were initially outside this sphere, but inside the sphere of $800$~au in radius\footnote{We note that it would be possible to use also an initialization sphere that extends beyond the adopted limit, that would also require to use longer initialization time and significantly larger number of objects. However, this would notably increases computational cost, with no obvious benefit for our work.}. The main characteristics of the three spheres described above are summarized in Table~\ref{tab:spheres}.

\begin{table}
	\caption{Summary of the adopted characteristics of observable, model and initialization sphere. See text for additional details.}
	\label{tab:spheres}
	\begin{tabular}{| l | l |} 
		\hline\hline
		observable  & largest object visible\\
		sphere - 12 au & at the edge under ideal conditions\\
		\hline
		model  & fastest object from the population\\
		sphere - 97 au & can reach observable sphere in\\
		 & 10 years simulation time\\\\
		 & all objects from the population leave this\\
		 & sphere in initialization time of 80 years\\
		 \hline
		 initialization & fastest object at the edge of this sphere\\
		 sphere - 800 au & can reach the model sphere in initialization\\
		 & time of 80 years\\\\
		 & Inside this sphere (and outside the model sphere) is \\
		 & assumed that the  gravitational focusing is negligible, \\
		 & and that distribution of ISOs is homogeneous \\
		 & and isotropic\\
		\hline\hline
	\end{tabular}
\end{table}

\item Assuming number density of 10 objects per au$^3$ for object larger than 10~m (described above), we generated $\approx21$ billion objects randomly distributed inside the initialization sphere, with isotropicaly distributed velocity vectors whose intensities follow the already mentioned normal distribution ($v_0=25$ km/s , $\sigma=5$ km/s). The Cartesian state vectors of these objects were then converted to the Keplerian orbital elements \citep{Bate_1971}, and their positions after 80 years were determined by solving hyperbolic Kepler equation (Eq.~\ref{eq:H}). Finally, objects located inside the model sphere ($\approx43$ million) were selected for further analysis.

Fig.~\ref{fig:number-density} shows the variation of the number density of the selected population, compared to the value in the initialization sphere where we assumed that the gravitational focusing is negligible, versus the distance from the Sun.
It can be seen that due to the gravitational focusing at $1$~au from the Sun the number density should be twice higher than the assumed value in the initialization sphere. On the other hand, one can notice that at the edge of the model sphere, the number density is already very close to the value in the initialisation sphere, which means that the assumption about homogeneous and isotropic population outside this sphere is valid.

\begin{figure}
	\includegraphics[width=\columnwidth]{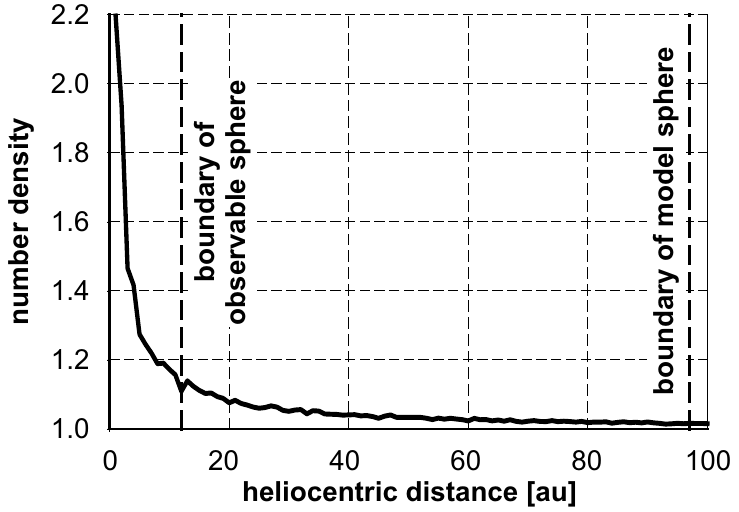}
    \caption{Variation of the number density of ISOs with heliocentric distance. The values on y-axis are normalized to the assumed value in the initialisation sphere, outside of the model sphere, which is unaffected by the gravitational focusing.}
    \label{fig:number-density}
\end{figure}

\begin{figure}
	\includegraphics[width=\columnwidth]{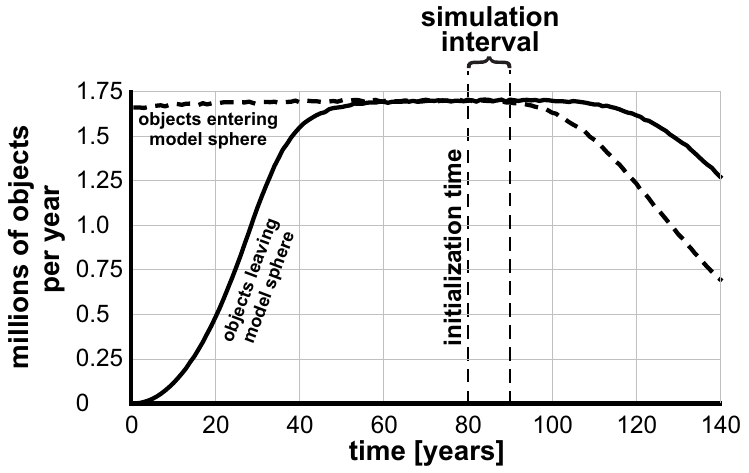}
    \caption{Number of objects that enter and leave the model sphere as a function of time. We stress that these numbers are roughly equal after the initialization time of 80 years.}
    \label{fig:in-out}
\end{figure}

Another important aspect to consider is how the number of objects that enter and exist from the model sphere evolves with time. As can be seen in Fig.~\ref{fig:in-out}, an initial flux towards the model sphere, of about 1.7 millions of objects per year, remains constant till about 85 years since the beginning of the simulation.\footnote{This drop in a number of objects entering the model sphere is simply a consequence of limited size of the initialization sphere. For our purpose here, the stable flux towards the model sphere is lasting long enough, but in principle it could be extended to any desirable time, by changing appropriately a size of the initialization sphere.} A number of objects exiting from the model sphere is initially zero, because the sphere is originally empty, and increases till it reaches the inward flux, that happens after 80 years. Therefore, after the initialization time interval of 80 years, the inward and outward flux are in balance, meaning that the population of objects inside the model sphere is in the steady-state.

\begin{figure}
	\includegraphics[width=\columnwidth]{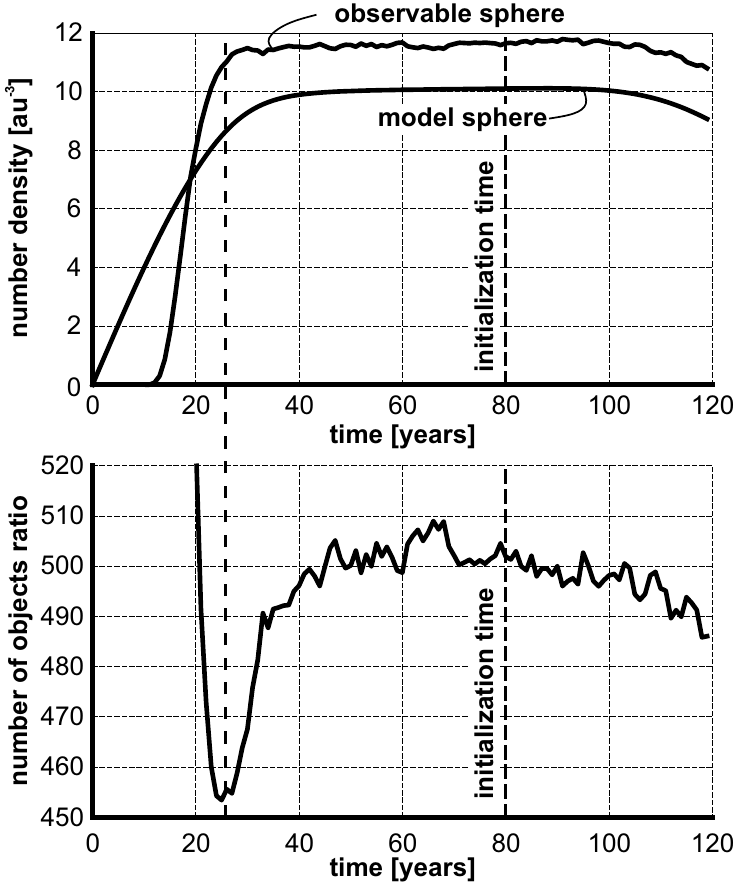}
    \caption{Time variations in the number density (upper panel) and in the ratio of the number of objects (bottom panel) inside the model and observable spheres.}
    \label{fig:model-obsrvations}
\end{figure}
Moreover, as Fig.~\ref{fig:model-obsrvations} shows, the ratio between the number of objects inside the model ($r=97$ au) and observational ($r=12$ au) spheres, is stable around the end of the initialization interval. At this point, the value of the ratio is about $500$, that is somewhat below the ratio of the volumes of the two spheres, due to the increased number density of objects closer to the Sun, as a result of the gravitational focusing. While the model sphere starts to fill immediately after the simulation begins, the observable sphere remains completely empty until the first objects manage to reach it, after more than 10 years. Since this point, as the consequence of the gravitational focusing, the observable sphere is filled faster than the model sphere, due to the larger number density of objects just outside the observable sphere than just outside the model sphere. 

Because of this, the observable sphere reaches an equilibrium number of objects a little earlier, about 30 years after the start of the simulation, leading to a rise in the number density ratio, until a stable value is reached.

Finally, in order to estimate a role of planetary perturbations, we also randomly selected a smaller sample of $1$ million objects, and propagated their trajectories using Bulirsch-Stoer algorithm as implemented in a public domain \emph{Mercury} software package \citep{1999MNRAS.304..793C}. The orbits of these objects were followed for $80$~yr, within the dynamical model that includes gravitational effects of the Sun and eight major planets. As expected, no statistically significant differences were noticed between this sample and the overall population which is propagated by means of hyperbolic Kepler equation.

\end{enumerate}

\subsubsection{Orbital distributions}

\begin{figure}
	\includegraphics[width=\columnwidth]{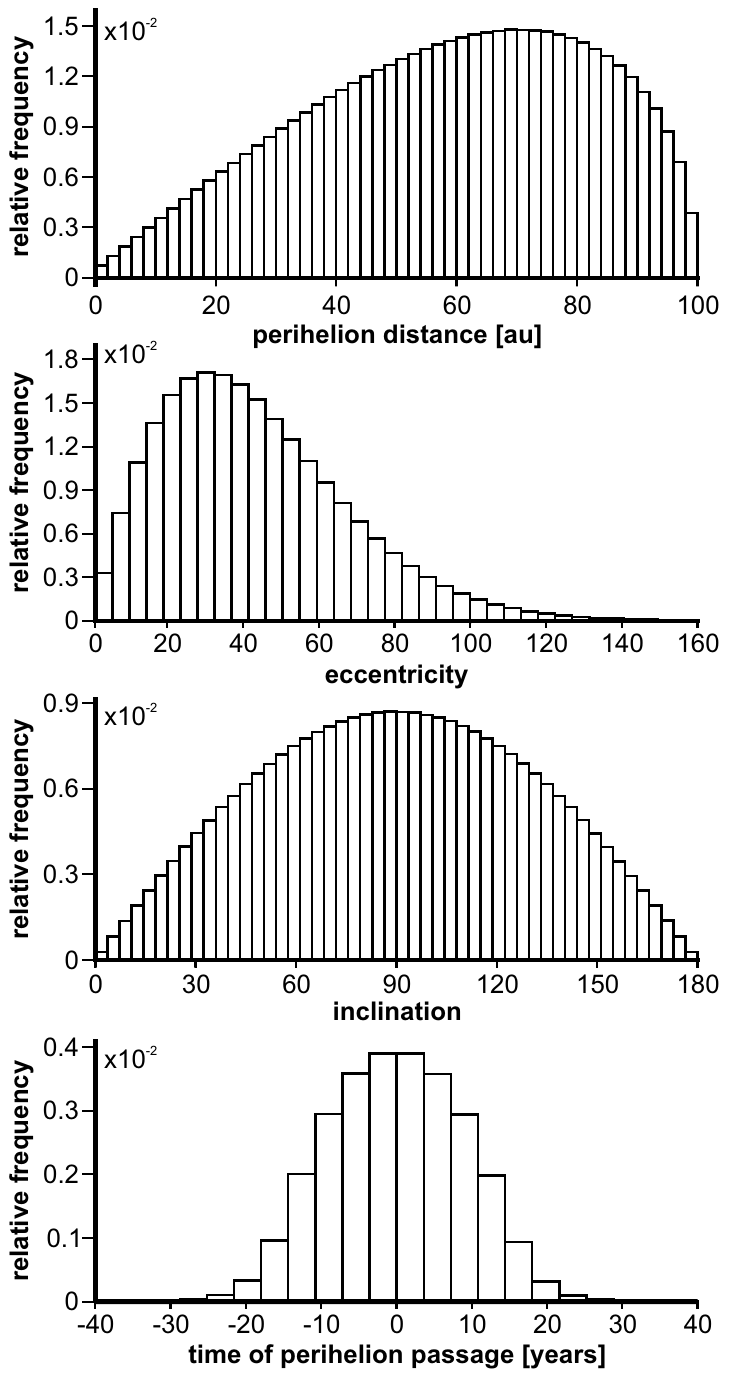}
    \caption{Distribution of orbital elements for the generated population of ISOs inside the model sphere, taken at the end of the initialization time.}
    \label{fig:ISO orbital elements}
\end{figure}

\begin{figure}
	\includegraphics[width=\columnwidth]{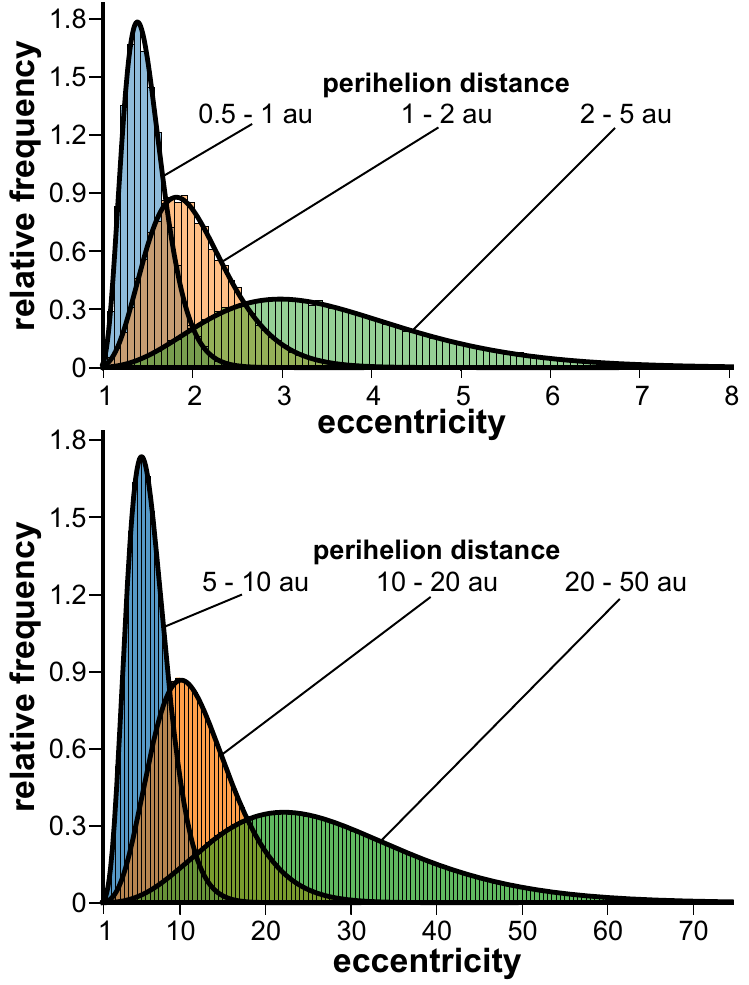}
    \caption{Orbital eccentricities of ISOs approximated with Gamma functions, for different ranges of perihelion distances. The two panels show normalized histograms for 6 samples of the generated population of ISOs. All histograms excellently follow Gamma distributions (shown as black lines). The samples are created according to perihelion distance, since it is directly related to eccentricity for a given $v_{\infty}$. The assumed normal distribution of $v_{\infty}$ ($v_0=25$ km/s, $\sigma=5$ km/s) results in Gamma-like distribution of eccentricities, with the parameters depending on the perihelion distances.}
    \label{fig:gamma_2D}
\end{figure}

\begin{figure}
	\includegraphics[width=\columnwidth]{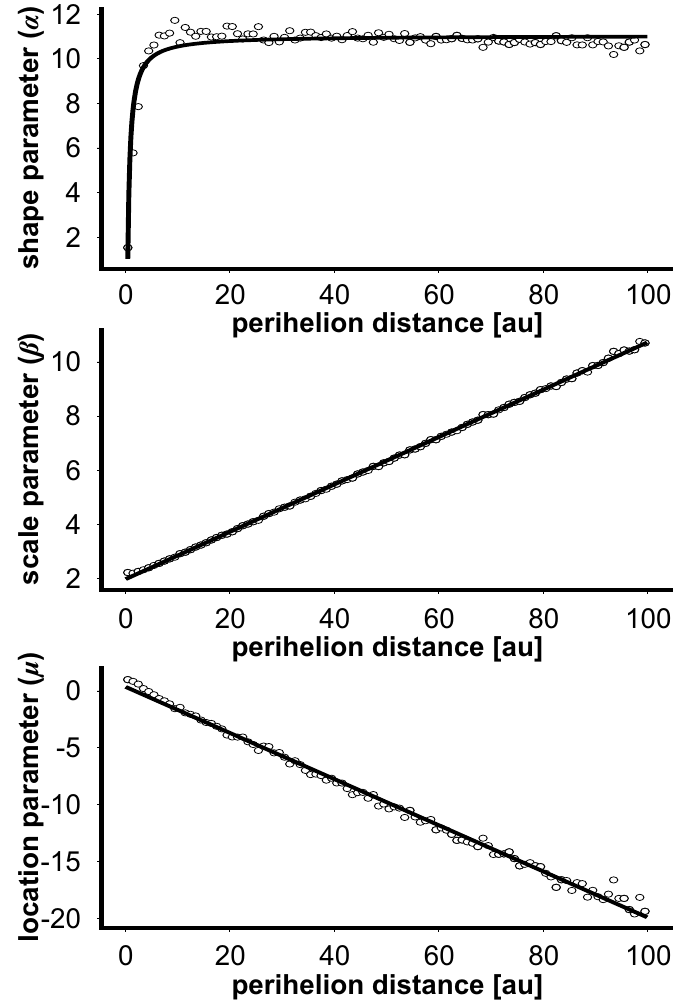}
    \caption{Dependence of the parameters of the fitted Gamma distributions on
    the perihelion distance. The dots present parameters obtained for consecutive intervals of perihelion distance of $1$~au, while the solid lines are their corresponding fits. The parameters are fitted by appropriate rational functions (for shape parameter - $\alpha$) and linear functions (for scale and location parameters - $\beta$ and $\mu$, respectively).}
    \label{fig:gamma_parameters}
\end{figure}

\begin{figure}
	\includegraphics[width=\columnwidth]{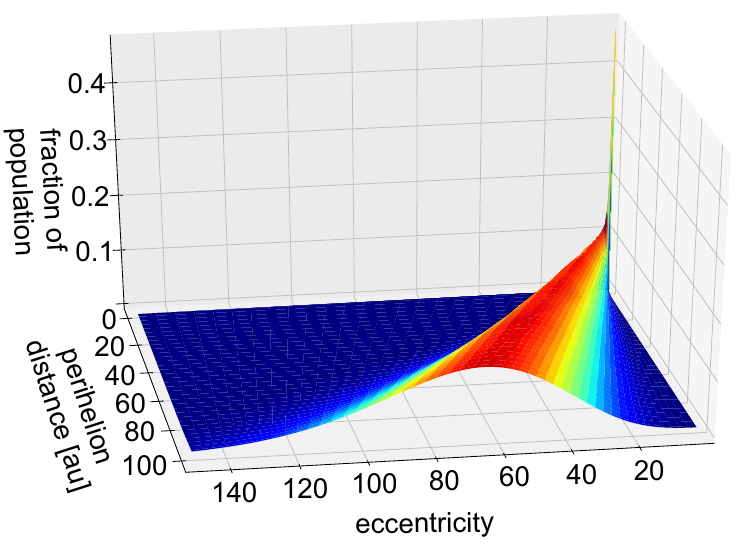}
    \caption{Bi-variate distribution of perihelion distances and eccentricities. The surface is obtained by generating Gamma distribution defined by Eq.~\ref{eq:gama_pdf}, using distribution's parameters defined by Eq.~\ref{eq:gama_parameters}.}
    \label{fig:bi-variate distribution}
\end{figure}

The distributions of the orbital elements of the resulting population of all objects within the model sphere is shown in Fig.~\ref{fig:ISO orbital elements}. The sinusoidal distribution of orbital inclinations is a consequence of the fact that the orbital normal vectors are randomly distributed over the sphere, which result in larger number of highly inclined orbits. Beside this, as expected, longitudes of nodes and arguments of perihelions are uniformly distributed.
Longitudes of nodes, arguments of perihelions and inclinations depend only to the initial positions and orientations of the velocity vectors, and are therefore expected to be mutually independent. On the other hand, perihelion distance and eccentricity are related through the equation \citep{kemble_2006}:
\begin{equation}
q=a+\sqrt{a^2+B_{dist}^2},
\label{eq:q(a)}
\end{equation}
where $a$ is semi-major axis, and $B_{dist}$ is distance, measured in the b-plane\footnote{The b-plane is defined to contain the focus of an idealized two-body trajectory that is assumed to be a hyperbola, and is perpendicular to the incoming asymptote of the hyperbola.}, between the trajectory defined by the initial velocity vector ($\vec{v}_{\infty}$) and the Sun. Taking into account Eq.~\ref{eq:e(q)}, a relation between perihelion distance and eccentricity can be written in the form:
\begin{equation}
q=B_{dist} \sqrt{{e-1} \over {e+1}}.
\label{eq:q(e)}
\end{equation}
{We noticed that distribution of orbital eccentricities, for sample of orbits inside certain range of perihelion distances, excellently follows Gamma distribution of the form}
\begin{equation}
f \left( e \right)=\frac{\left(e-\mu \right)^{\alpha-1}}{\beta^{\alpha} \Gamma \left(\alpha \right)} \exp{\left(-\frac{e-\mu}{\beta}\right)},
\label{eq:gama_pdf}
\end{equation}
where $e$ is orbital eccentricity, and $\alpha$, $\beta$ and $\mu$ are shape, scale and location parameters of the Gamma distribution, respectively. The approximations for different ranges of perihelion distances are shown in Fig.~\ref{fig:gamma_2D}.
Based on our numerical experiments (see Fig.~\ref{fig:gamma_parameters}), we found that the parameters of these Gamma distributions are in simple relations with the perihelion distance, as given by the formulas:
\begin{equation}
\begin{aligned}
\alpha &= 11.042-4.966 / q \\
\beta &= 0.087q-0.022 \\
\gamma &= -0.203q+0.377
\label{eq:gama_parameters}
\end{aligned}
\end{equation}
\noindent The Eqs.~\ref{eq:gama_pdf} and~\ref{eq:gama_parameters} define the analytical expression for the bi-variate distribution of perihelion distance and eccentricity of our generated population. The obtained distribution is presented in Fig.~\ref{fig:bi-variate distribution}.

The analytic expressions for orbital distributions allow direct sampling of orbits from these distributions, without further need for the previously described complex algorithm. Analytical models of the populations are very useful in estimating the observational constraints and selection effects of the populations. Similar expressions are already determined for some populations in the Solar System such as the model of Centaur objects \citep{1997Icar..127..494J} incorporated in the comprehensive model of the Solar System (S3M)\footnote{This model is used to evaluate the performance of Pan-STARRS survey in discovering objects from various populations of the Solar System, including also the interstellar comets.} by \citet{2011PASP..123..423G}.

\section{Analysis and Discussion}

\subsection{Orbit and size distribution of observable objects}

To analyze the objects from our ISOs population, which could be detected by the VRO, we conducted the simulation in which we calculated geocentric coordinates, solar elongation and apparent brightness of the objects for every hour inside the simulation period of ten years. In order to identify potentially observable objects, we adopted detectability conditions (summarized in Table~\ref{tab:observational constraints}) based on the nominal characteristics of the so-called Wide, Fast, Deep (WFD) observational proposal of the LSST \citep{2017arXiv170804058L,2016IAUS..318..282J}. Finally, we identified all the objects that satisfy the detectability conditions in at least one simulation time step.
\begin{table}
	\caption{Detection constraints based on the nominal characteristics of the LSST observational program. An object is considered as observable if in at least one time step within the 10 years of the simulation it satisfies the constraints given in this table. The limitations guarantee that the object is bright enough while located in the appropriate part of the sky to be observed by  LSST. We emphasize that a part of the sky around the Galactic plane is excluded.}
	\label{tab:observational constraints}
	\begin{tabular}{| l | c |} 
		\hline\hline
		Apparent visual magnitude& $m<24.5$\\
		\hline
		Declination & $-65^{\circ} < \delta <5^{\circ}$\\
		\hline
		Elongation & $>60^{\circ}$\\
		\hline
		Galactic coordinates limits& $\left| b\right|=\left(1-l/90^{\circ}\right)\times10^{\circ}$ \\
		 & $0^{\circ}<l<90^{\circ}$\\
		 & $270^{\circ}<l<360^{\circ}$\\
		 & where $b$ and $l$ are galactic\\
		 & latitude and longitude, respectively\\
		\hline\hline
	\end{tabular}
\end{table}

Clearly, the restrictions given in Table~\ref{tab:observational constraints} are far from being sufficient to make any object detected, and especially identified as unknown. A probability that the object will be detected depends on many other factors, such as seeing conditions, effects of the Moon, detection and trailing losses, observing cadence,  etc. Moreover, a chance that the detected object will be identified as interesting for follow-up, that would lead to its orbit determination and classification as interstellar, depends on complex set of parameters included in the Minor Planet Center's so-called digest score \citep{2019PASP..131f4501K}. However, the constraints given in Table \ref{tab:observational constraints}, are for sure the necessary conditions that any object potentially detectable by LSST must satisfy.

To take into account only objects that may satisfy these conditions, from the previously described global ISOs population in the model sphere ($\approx43$ million), we selected only those ($\approx380 000$) which were initially inside the observable sphere, or appear inside this sphere during the simulation time of 10 years, based on the hyperbolic Kepler equation. This does not guarantee that these objects will reach the defined limiting apparent magnitude, and/or to appear in the appropriate part of sky to be observed. It only implies that no other objects from our synthetic population can be observed during the VRO operational period.

\begin{figure}
	\includegraphics[width=\columnwidth]{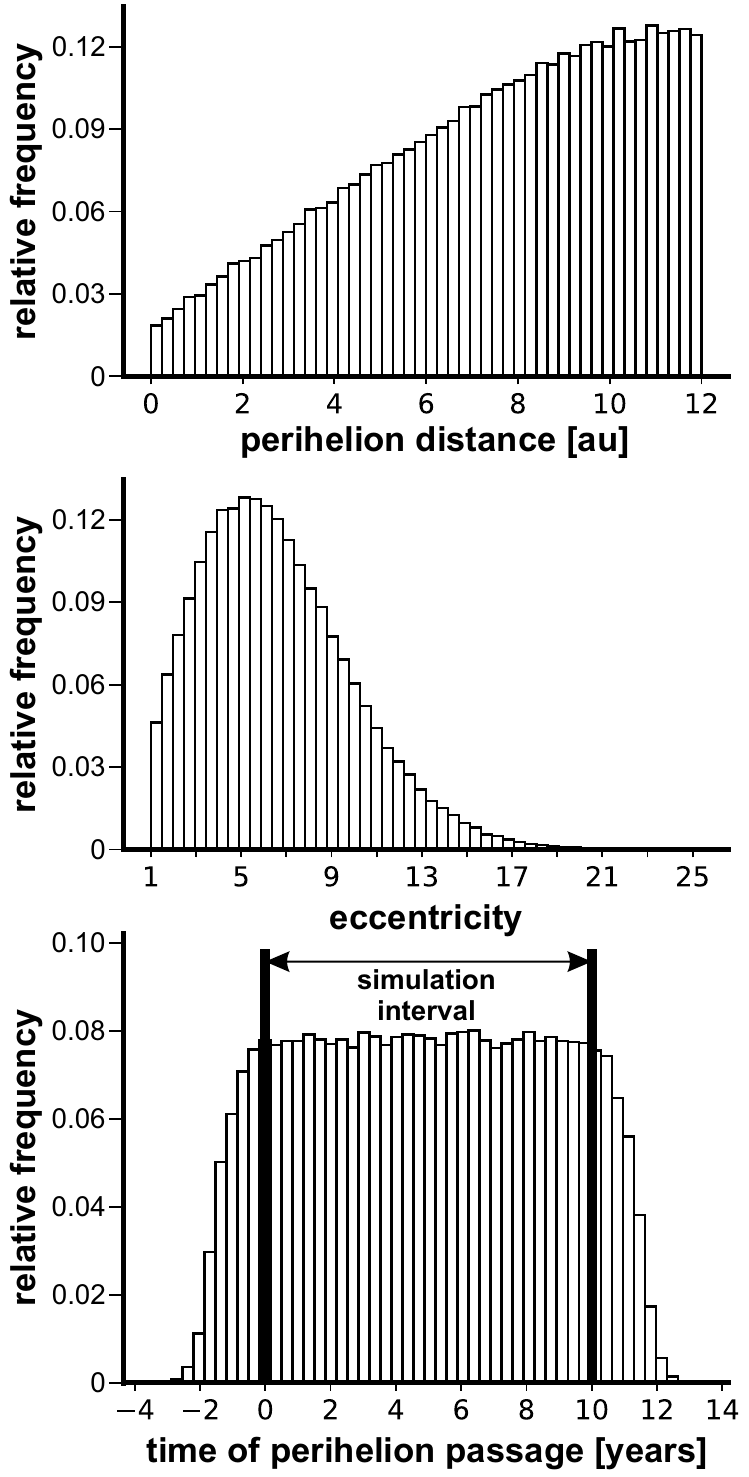}
    \caption{Distribution of orbital elements of potentially detectable objects. The graphs present normalized histograms of orbital elements for objects which are located inside the observable sphere ($r=12$ au) at the beginning of the simulation, or appear inside it during the simulation period.}
    \label{fig:detectable_objects}
\end{figure}

In Fig.~\ref{fig:detectable_objects}, orbital elements of potentially observable objects are shown.
In the bottom panel of this figure one can see that the distribution of the times of perihelion passages is uniform over the simulation period, which, combined with steadiness of distributions of other orbital elements, means that the generated population is time independent during this period.

\begin{figure}
	\includegraphics[width=\columnwidth]{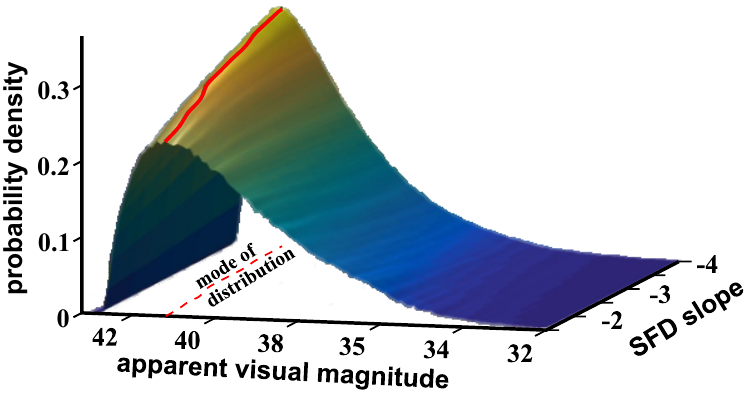}
    \caption{Graph shows the distribution of the apparent magnitudes of the whole synthetic population, at an arbitrary epoch, for different SFD slopes. The mode of the distribution is highlighted to emphasize the apparent faintness of ISOs.}
    \label{fig:apparent magnitudes}
\end{figure}

\begin{figure}
	\includegraphics[width=\columnwidth]{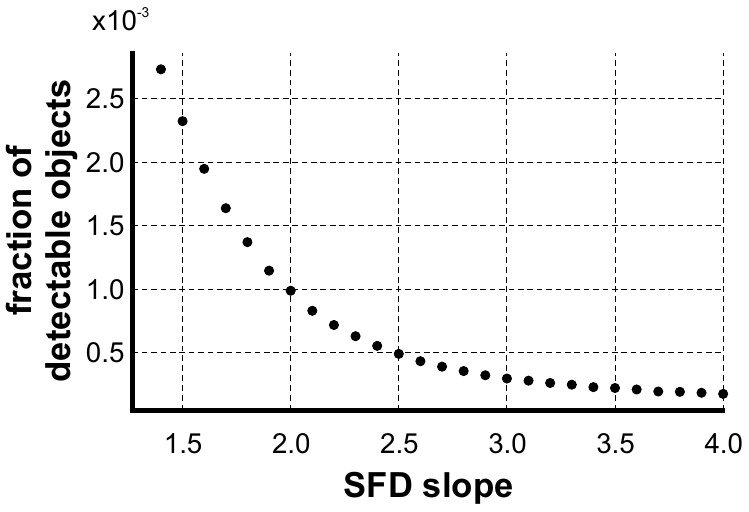}
    \caption{Dependence of the fraction of detectable objects on the population's SFD slope.}
    \label{fig:detectable_frequency}
\end{figure}

Taking into account the size range of the generated population, the distribution of orbital elements (primarily the perihelion distance), and the assumed albedo of $0.04$, these objects are expected to be very faint, that is the main difficulty for their detection. Fig.~\ref{fig:apparent magnitudes} shows the distribution of the apparent magnitudes for different SFD slopes, at an arbitrary epoch. 
It is obvious that regardless of the SFD slope, the mode of this distribution, is at any epoch far beyond capabilities of the current and planned wide-field surveys. Only objects from the far tail of the distribution may be detected. This is further illustrated in Fig.~\ref{fig:detectable_frequency}, where one can see the frequency of the objects, among the whole population, which satisfy the conditions given in Table~\ref{tab:observational constraints}, and may possibly be detected.
The results show that for steeper SFD slopes, at best, only one in several thousands objects should be expected to fulfill the minimum detectability criteria.

\begin{figure}
	\includegraphics[width=\columnwidth]{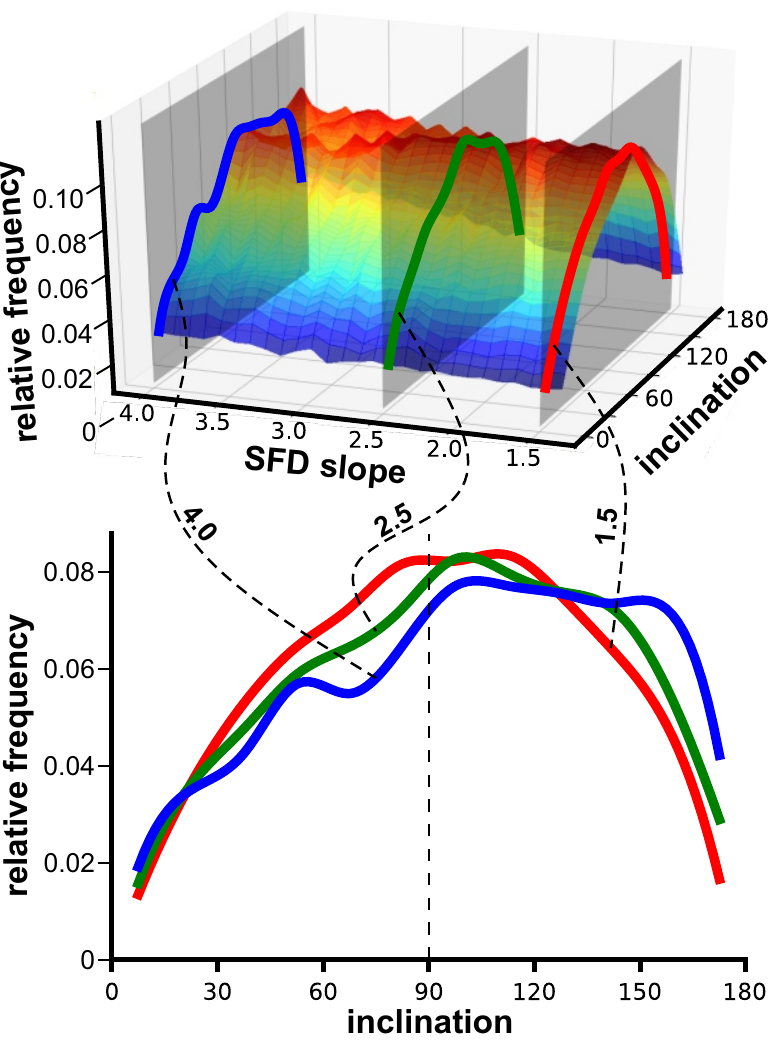}
    \caption{Distributions of orbital inclinations. The upper panel shows the distribution of orbital inclinations of the observable objects depending on the SFD slope of the underlying true population. The lower panel shows the inclination distributions (normalized histograms and their interpolated curves) for the three selected values of SFD slope ($1.5$, $2.5$ and $4$), which are also highlighted in the upper panel. It can be seen that, as the SFD slope increases, the maximum of the distribution shifts toward larger inclinations.}
    \label{fig:inclinatios}
\end{figure}

Analyzing the orbital elements of the detectable objects (those which happen to satisfy constraints given in Table \ref{tab:observational constraints} during the simulation interval) we noticed that the most prominent feature is asymmetric distribution of their orbital inclinations\footnote{We recall here that \citet{2017AJ....153..133E} have also noticed similar fact, but attributed this to the digest score flag that may favor the retrograde orbits. However, although the digest score may play a role in the distribution of detected ISOs, there must be other reason as well, because in our study we did not consider
efficiency of the Moving Object Processing System.}, reflected through a larger number of retrograde ($i>90^{\circ}$) than direct orbits ($i<90^{\circ}$), as shown in Fig.~\ref{fig:inclinatios}.

\begin{figure}
	\includegraphics[width=\columnwidth]{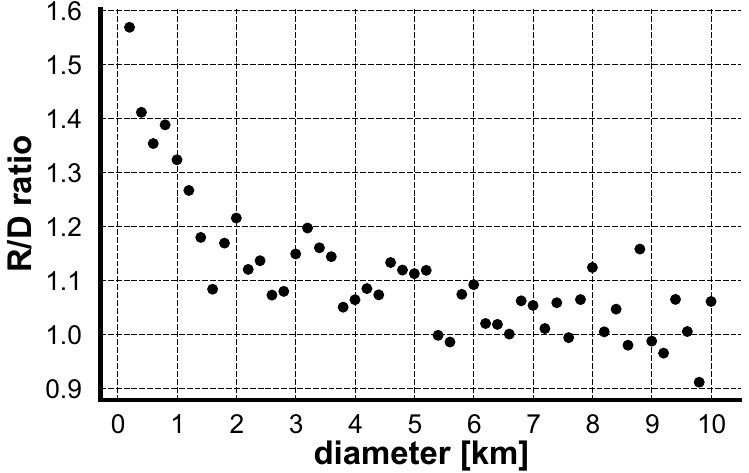}
    \caption{R/D ratio for detectable objects of different sizes, binned by $200$~m.}
    \label{fig:D-depandance}
\end{figure}

In order to identify which factors affect this asymmetry, we examined how the distribution of orbital inclinations depends on possibly relevant parameters. More in particular, to explore if the ratio of the numbers of retrograde and direct objects (R/D ratio) is size dependant, we analyzed a set of detectable objects for 27 different SFD slopes (from $1.4$ to $4$). This set of objects was divided in subsets based on diameters, with a step of $200$~m, and the R/D ratio for each of these subsets was calculated. The obtained results are shown in Fig.~\ref{fig:D-depandance}. In this figure, the data for all the SFD slopes are shown together. This is because we would like to highlight how the R/D ratio changes for different sizes of objects, and plotting all the objects together provides a better statistical sample.

\begin{figure}
	\includegraphics[width=\columnwidth]{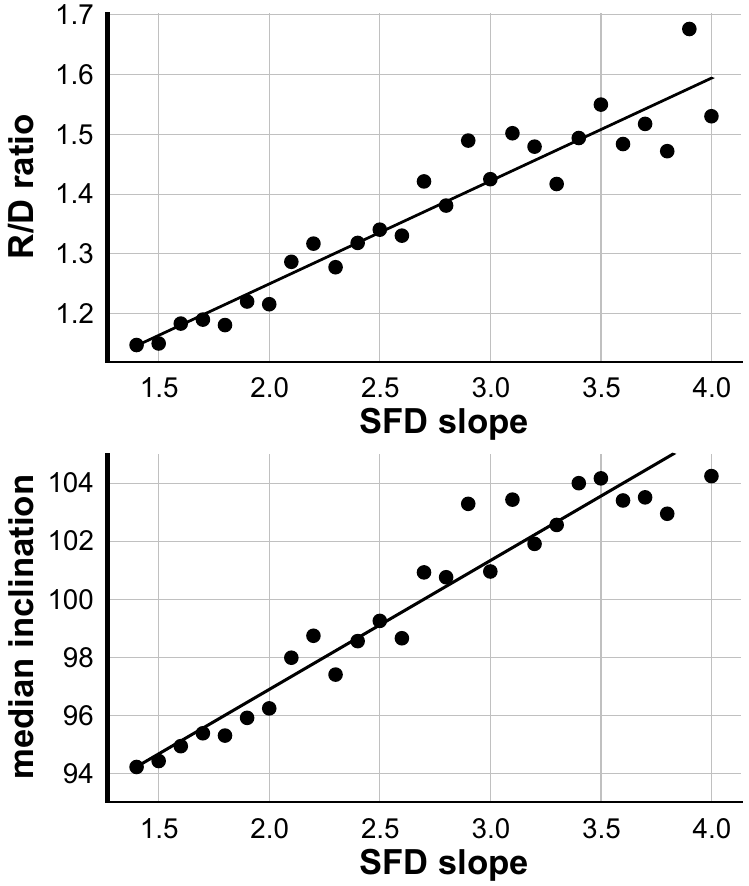}
    \caption{Dependence of R/D ratio (upper panel) and median inclination (lower panel) of detectable objects on the SFD slope of the underlying true population.}
    \label{fig: correlation}
\end{figure}

 It is noticeable that, for objects of several kilometers in diameter, there is almost no difference, while there is a large asymmetry for sub-kilometer objects, with the latter group making 3 out of 5 to be retrograde. The consequence of this phenomenon is that the R/D ratio, as well as the median inclination, is correlated with the SFD slope of the true population, because steeper SFDs have more smaller objects that, in turn, results in a higher R/D ratio. To clearly illustrate this fact, in Fig.~\ref{fig: correlation} we plot the median inclination and the R/D ratio as a function of the SFD slope. The results shown in this figure are basically the same ones as those shown in Fig.~\ref{fig:D-depandance}, but this time separated based on the slopes, instead of the diameters. As can be seen in Fig.~\ref{fig: correlation}, both parameters depend on the SFD slope. This fact could allow preliminary estimation of the SFD slope of the true population, based on the orbital inclinations of known population, once a sufficient number of objects have been discovered.

\begin{figure}
	\includegraphics[width=\columnwidth]{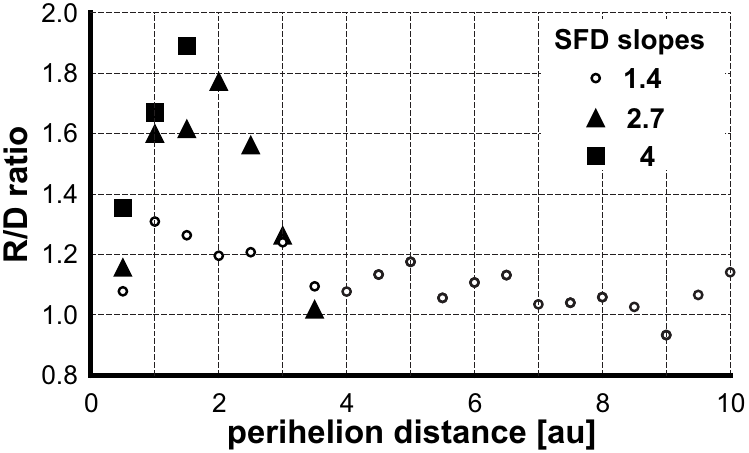}
    \caption{R/D ratio for detectable objects for different perihelion distances, binned by $0.5$~au. We plot only data for bins containing at least 100 objects.}
    \label{fig:perihelion}
\end{figure}

Beside the SFD slope, we also examined the dependence of the asymmetry on other parameters, and found a potentially interesting dependence of the R/D ratio on the perihelion distance. Similar to the analysis shown in Fig.~\ref{fig:D-depandance}, from the set of all detectable objects we took the subsets of objects inside perihelion distance limits, with a step of $0.5$~au, and calculated R/D ratios of these subsets. The obtained results for three different slopes of the SFD are shown in Fig.~\ref{fig:perihelion}. 
It should be noticed that the objects with smaller perihelion distances are the most strongly influenced, with maximum around 1-2~au. In addition, this dependence is more pronounced for steeper SFD slopes. 

\subsection{The role of Holetschek's effect}

The question is what causes that the majority of the detectable ISOs have retrograde orbits, and through which mechanism(s) the R/D asymmetry is related to the SFD and perihelion distances? To answer these questions we turn our attention to long period comets, that could be affected by the same orbital biases as interstellar objects. Still, we need to keep in mind a fact that the LPCs are periodic (returning objects), while ISOs are not. Moreover, in this work we are considering only asteroid-like ISOs, so brightness increase due to the activity is not taken into account here. Nevertheless, some results about the LPCs seem to be useful to explain some of our findings.

An excess of the retrograde orbits has been noticed among the LPCs \citep[see e.g.][]{1967AJ.....72.1002E,1981MNRAS.197..265F,1991Natur.352..506M,2016AJ....152..103S}, although more recent results suggest that it may not be so pronounced \citep{2019AJ....157..181V}. There is still no consensus if this phenomenon is a result of observational selection effect, or there is a real asymmetry in the population, as a consequence of some dynamical mechanism. For instance, on one side, \citet{1972A&A....20..205Y} argued that direct comets are exposed to stronger action of planetary perturbations, leading to their faster dynamical evolution, and consequently elimination from the Solar System. This claim seems however to be disputed by findings of \citet{1981MNRAS.197..265F}, who found the same asymmetry among the orbits of young comets, that have not had time to evolve due to the planetary perturbations, indicating that the pattern cannot be explained by an aging effect. Also, \citet{1991Natur.352..506M} suggested dynamical explanation for the excess of observed comets in retrograde orbits. These authors proposed it could be due to enhanced volatility of retrograde comets, as a result of more energetic collisions with direct meteoroids, comparing to direct comets. The latter explanation however can not be applied to our results for ISOs, because it involves cometary activity.

\begin{figure}
	\includegraphics[width=\columnwidth]{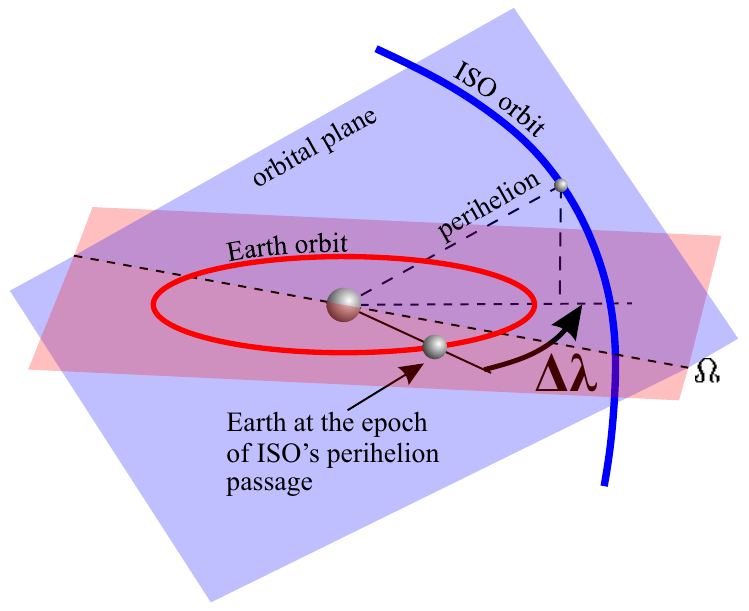}
    \caption{Illustration of the orbital and position constellation related to Holetschek's effect. Quantity $\Delta \lambda$ is the difference between the heliocentric ecliptic longitudes of an interstellar object and the Earth, at the epoch of the object's perihelion passage.}
    \label{fig:longitudes}
\end{figure}

An alternative explanation for a possible excess of retrograde objects among known LPCs is Holetschek's effect. According to this effect, objects which reach perihelion on the side of the Sun opposite to the position of the Earth are less likely to be discovered \citep{Holetschek,1967AJ.....72..716E,1983MNRAS.204...23H,2002MNRAS.335..641H}. This is because in this configuration objects are both, too close to the Sun (small elongation), and further away from the Earth (fainter). Therefore, a probability for an object to be discovered depends on the difference ($\Delta \lambda$) between the heliocentric longitude of the Earth and that of the object at the time of the perihelion passage of this object. This is illustrated in Fig.~\ref{fig:longitudes}.

This effect is more important for direct than for retrograde orbits. The qualitative explanation is as follows: after perihelion, the Earth and the retrograde object are, on the average, moving towards each other, reducing quickly $\Delta \lambda$ angle. Therefore, although object's heliocentric distance is somewhat increasing, its geocentric distance is dropping comparatively quickly, which in turn allow some of the retrograde objects to be discovered after their perihelion passage. On contrary, the objects on direct orbits are, after perihelion, moving typically away from the Earth, leaving almost no chance to be discovered. This means that retrograde objects which are not in observable position when they are at perihelion, have much better chance to take a more observable position, before moving too far from the Sun. 

\begin{figure}
	\includegraphics[width=\columnwidth]{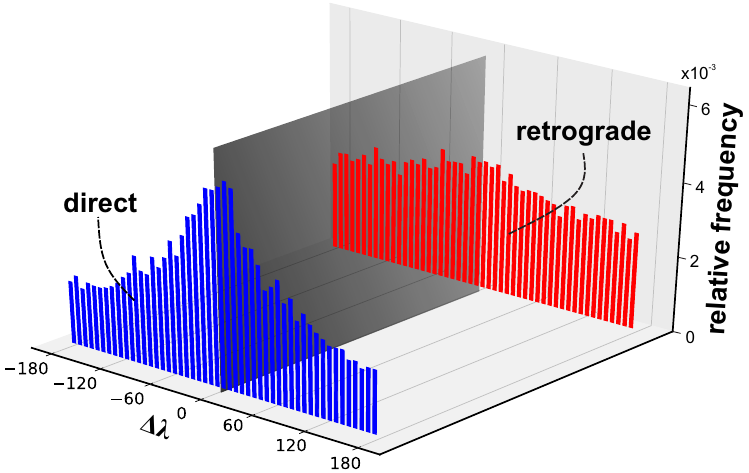}
    \caption{Holetschek's effect for direct and retrograde objects. This figure shows normalized histograms of $\Delta \lambda$ of the detectable objects, separately for those on direct ($i<90^{\circ}$) and retrograde ($i>90^{\circ}$) orbits. The vertical plane denotes objects which pass through their perihelions while they are on the same heliocentric ecliptic longitude as the Earth.}
    \label{fig:Holetschek 1}
\end{figure}

Having in mind that our population of ISOs is generated as symmetric, there is no doubt that the asymmetry of orbital inclinations is due to a selection effect. We examined the distribution of the angle $\Delta \lambda$ for the observable ISOs and found clear distinction between direct and retrograde orbits, as shown in Fig.~\ref{fig:Holetschek 1}. For direct orbits there is a strong concentration around $\Delta \lambda=0$, while for the retrograde objects this distribution is almost uniform. This is a strong indication that Holetschek's effect is responsible for the R/D asymmetry in our data.

\begin{figure}
	\includegraphics[width=\columnwidth]{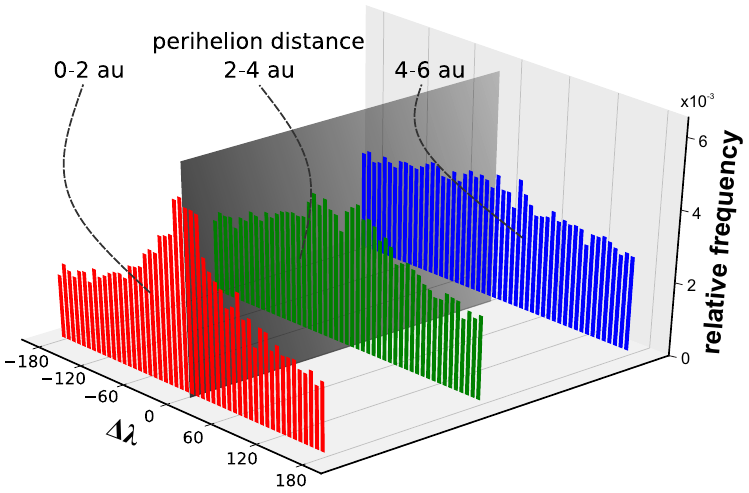}
    \caption{Holetschek's effect for different ranges of perihelion distances. The normalized histograms of $\Delta \lambda$ of the detectable objects, for three ranges of perihelion distances. The vertical plane denotes objects which pass through their perihelions while they are on the same ecliptic longitude as the Earth. Data shown in the plot include both, direct and retrograde orbits.}
    \label{fig:Holetschek 2}
\end{figure}

\citet{1975BAICz..26...92K} found Holetschek's effect to be strongly dependent on the perihelion distance \citep[see also][]{Holetschek}. Briefly, the author found that between $q=0.5$ and $q=2$~au, the distribution of orbits is strongly biased, and the effect reaches its maximum for perihelion distances around $1$~au. On the other hand, for $q\leq0.5$ Holetschek's effect should be negligible, while beyond $q\approx2$~au it almost disappears. Therefore, if the observed asymmetry of the R/D ratio is a consequence of Holetschek's effect, our data should exhibit similar patterns. In this respect, we note that the results shown in Fig.~\ref{fig:perihelion} already point out in this direction. Still, to further clarify this we analyzed the distribution of $\Delta \lambda$ of the detectable objects, for three different ranges of perihelion distances. Our data shown in Fig.~\ref{fig:Holetschek 2} are exhibiting a very similar pattern as the one found by \citet{1975BAICz..26...92K}. For $q\in[0,2]$~au, there is a peak in the distribution of detectable objects in terms of $\Delta \lambda$. Some deviation from random distribution is also visible for $q\in[2,4]$~au, while for $q\in[4,6]$~au the distribution is uniform.

The fact that Holetschek's effect is more important for direct than for retrograde orbits, and that it mainly affects the orbits with $q\in[0.5,2.5]$~au, fully explains the results presented in Fig.~\ref{fig:perihelion}. The excess of retrograde objects is the most pronounced for orbits with {$q\approx 1.5$~au}, the ones strongly affected by observational bias caused by the aforementioned effect. Taken together, these results clearly indicate that Holetschek's effect is responsible for the excess of the retrograde orbits among the interstellar objects observable by the VRO. Having saying that, we do not exclude that other observational selection effect also contribute to the excess of retrograde orbits, but to a somewhat lesser extent \citep[see e.g.][for a review on other observational biases]{1975BAICz..26...92K,2002MNRAS.335..641H}.

In addition, as noted by \citet{1983MNRAS.204...23H}, the effect is stronger for smaller than for larger objects, because the larger objects are on average brighter, and visible at larger heliocentric and geocentric distances. For this reason, the observational window of larger objects is longer, and therefore they are less sensitive to the effect. This practically means that Holetschek's effect is size dependent. Our results shown in Fig.~\ref{fig:D-depandance} fully support this fact. This is also the reason for the dependence of the R/D ratio on the SFD slope. A stepper size-distribution implies more small objects, which makes the considered population more affected by Holetschek's effect. This concept explains the dependence of the R/D ratio and the median inclination on the SFD slope shown in Fig.~\ref{fig: correlation}. 

\subsection{On some limitations of our approach}

We would like here to discuss some limitations of the obtained results and prospects for future work.

As already mentioned in Section~\ref{ss:num-density}, a single power-law approximation of the size-frequency distribution used here, may not actually be the best option. For some
populations of small objects in the Solar System, as for instance in the Kuiper belt, it is
well known that the slope becomes significantly shallower at smaller sizes \citep[e.g.][]{2019Sci...363..955S}. This would imply that there should be comparatively less smaller objects than in our simulation. As our analysis show that the R/D ratio is less significant for larger objects, the observed excess of retrograde orbits would be somewhat less pronounced in population described at smaller sizes with the shallower SFD slope.
Though we think our overall conclusions would be still valid, the results would be definitely different, and this deserve to be studied in future work.

Another limitation of the results presented here stems from the fact that a probability to detect an ISO depends on several factors that we did not considered here. These factors include seeing conditions, effects of the Moon, detection and trailing losses, observing cadence, digest score etc. For instance, \citet{2017AJ....153..133E} have also noticed an excess of retrograde objects in their data, but attributed this to the digest score flag that may favor the retrograde orbits. On the other hand, we found here that the excess of retrograde orbits exists even without the efficiency of the Moving Object Processing System
taken into account. Therefore, it would be important to investigate what is the R/D ratio when all the factors are taken into account simultaneously.

Finally, we considered here only asteroid-like interstellar objects, and neglected any cometary activity. To calculate apparent magnitudes we assumed simplified linear darkening function of the phase angle, and neglected any possible brightening in comet-like objects. 

Recently, \citet{2019AJ....157..162H} found that comet C/2010 U3 (Boattini) was active at a new record heliocentric distance of $25.8$~au. The second most distant activity is observed in comet C/2017 K2, found to be active at $23.75$~au \citep{2017ApJ...849L...8M}. CO-driven comets activity at large heliocentric distances has been also predicted by some models. There are several possible mechanisms for activity at these large distances. Sublimation rate is a nonlinear function of temperature, and can occur at low rates at large distances. The sublimation temperatures of the most abundant ices that can drive activity, $CO$, $CO_2$, and $H_{2}O$, are $25$~K, $80$~K, and $160$~K, respectively. The distance at which surface-ice sublimation becomes effective at driving comet activity is when the gas flow lifts sufficient dust from the surface to be detected from Earth. For water this is within the distance of Jupiter; for $CO_2$, this is at the distance between Saturn and Uranus; and for $CO$, it is at distances within the Kuiper Belt \citep[][see also \citet{2019AJ....157...65J}]{2009Icar..201..719M}, or even up to heliocentric distances of $85$~au \citep{2020AA....in..press}.

Having in mind that our estimated observable sphere has radius of $12$~au, 
a cometary-like activity might occur outside this sphere, and a comet-like ISO
may be potentially observable at larger heliocentric distances. This would affect
our results to some extent, because a larger observable sphere should be used,
that would also imply larger model and initialization spheres. Still, for vast majority
of comets, brightening should occur only at smaller heliocentric distances \citep{2009Icar..201..719M}. The total brightness of a comet is the sum of the brightness of the comet nucleus and the brightness of the coma. \citet{1999A&A...352..327F} found that beyond $\sim5$~au contribution of the coma brightness tends to zero, and beyond this distance the comet brightness depends on the heliocentric distance in a way very similar to asteroid-like objects. Therefore, the size of the observable sphere that we used here would be reasonably appropriate also for comet-like ISOs. However, due to the increased brightness within $\sim5$~au, cometary-like objects should be on average discovered at somewhat larger heliocentric distances. As a result, Holetschek's effect would be less important for these objects, and this would affect to some degree the estimated R/D ratio of observable ISOs. Finally, the results regarding the orbital distribution of observable objects should not be significantly affected. Anyway, some caution is need here, and we underline that our results are strictly speaking valid only for asteroid-like ISOs.

Let us also note here that it is suggested by \citet{2012MNRAS.423.1674F} that Holetschek's effect should be less relevant for modern sky-surveys. The reason why we see the consequences of this effect in our simulations, might be because we work with asteroid-like objects, or due to the fact that ISOs are passing only once through the observable sphere. However, as the LSST survey is planned to operate for elongations larger than $60\circ$ we believe the observed excess of retrograde objects is still mainly due to Holetschek's effect.

\section{Summary and Conclusions}

We study the distribution of the orbital elements of the generated synthetic population of the interstellar objects, specifically focusing on those observable by the VRO, based on the nominal characteristics of this survey. While a several other authors performed a similar investigation, they focused on the expected number of detectable ISOs, rather than on their orbital characteristics. 

Our main conclusions can be summarized in the following:
\begin{itemize}

\item The gravitational focusing should not affect significantly the orbital distribution
of the observable ISOs. At 1~au distance from the Sun only twice as many objects as in the interstellar 
space should be expected, but outside of the model sphere, the number density is very close 
to the assumed value in the initialisation sphere, meaning that the assumption about homogeneous and isotropic population outside the model sphere is valid.

\item The perturbation by the planets do not produce any significant effect on the orbital distribution of ISOs.

\item {We found that the distribution of orbital eccentricities can be very well approximated 
with Gamma distributions, whose parameters are functions of perihelion distance.
These analytical expressions for orbital distributions allow simple direct sampling 
of objects without the need for applying complex technique described above.}

\item Among the potentially observable ISOs, there is an asymmetry in the distribution 
of orbital inclinations, with an overabundance of retrograde objects.

\item The excess is the result of Holetschek's effect which is already suggested to be
responsible for the oversupply of retrograde objects among the observed long-periodic comets. Holetschek's effect depends on the objects' sizes and their perihelion distances. 

\item The excess of retrograde objects depends on objects' sizes and perihelion distances. This should allow estimation 
of the SFD of the underlying true population based on R/D ratio and median inclination of the
discovered population.

\end{itemize}

\section*{Acknowledgements}
\addcontentsline{toc}{section}{Acknowledgements}
We sincerely thank the reviewer for constructive criticisms and valuable comments, which were of great help in revising the manuscript.
The authors acknowledge financial support from the Ministry of Education,
Science and Technological Development of the Republic of Serbia through the project ON176011 "Dynamics and kinematics of celestial bodies and systems".

\bibliographystyle{mnras}
\bibliography{references} 

\bsp	
\label{lastpage}
\end{document}